\documentclass[%
reprint,
superscriptaddress,
amsmath,amssymb,
aps,
prl,
]{revtex4-1}

\usepackage[english]{babel}
\usepackage{graphicx}
\usepackage{color}
\usepackage{dcolumn}
\usepackage{bm}

\def\etal{{\textit{et al. }}}

\begin{document}
	

	\title{Impact of a minority relativistic electron tail interacting with a thermal plasma containing high-atomic-number impurities}
	
	\author{Nathan A. Garland }
	\affiliation{Los Alamos National Laboratory, Los Alamos, NM 87545, USA}
	
	\author{Hyun-Kyung Chung}
	\affiliation{National Fusion Research Institute (NFRI),
		169-148 Gwahak-ro, Yuseong-gu, Daejeon 34133, Korea}
	
	\author{Christopher J. Fontes}
	\affiliation{Los Alamos National
		Laboratory, Los Alamos, NM 87545, USA}
	
	\author{Mark C. Zammit}
	\affiliation{Los Alamos National Laboratory, Los Alamos, NM
		87545, USA}
	
	\author{James Colgan}
	\affiliation{Los Alamos
		National Laboratory, Los Alamos, NM 87545, USA}%
	
	\author{Todd Elder}
	\affiliation{Los Alamos
		National Laboratory, Los Alamos, NM 87545, USA}
	\affiliation{Columbia University, New York, NY 10027, USA}
	
	\author{Christopher J. McDevitt}
	\affiliation{Los Alamos
		National Laboratory, Los Alamos, NM 87545, USA}%
	
	\author{Timothy M. Wildey}
	\affiliation{Sandia National
		Laboratories, Albuquerque, NM 87185, USA}
	
	\author{Xian-Zhu Tang}
	\affiliation{Los Alamos National Laboratory, Los
		Alamos, NM 87545, USA}

	
	
	\date{\today}
	
	\begin{abstract}
		A minority relativistic electron component can arise
                in both laboratory and naturally-occurring plasmas.
                In the presence of high-atomic-number ion species, the
                ion charge state distribution at low bulk electron
                temperature can be dominated by relativistic
                electrons, even though their density is orders of
                magnitude lower. This is due to the relativistic
                enhancement of the collisional excitation and
                ionization cross sections. The resulting charge state
                effect can dramatically impact the radiative power
                loss rate and the related Bethe stopping power of
                relativistic electrons in a dilute plasma.\\(Approved for release under LA-UR-19-28749)
		
	\end{abstract}
	
	\maketitle
	
	
	
	Major disruptions must be adequately mitigated in a tokamak
        for a viable fusion power reactor, with the current approach
        for ITER's Disruption Mitigation System (DMS) based on
        injecting large quantities of high-atomic-number impurities
        such as argon or
        neon~\cite{lehnen2018,lehnenjournalofnuclearmaterials2015}.
        This can serve the dual purpose of thermal quench mitigation
        by spreading the plasma power load over the reactor first wall
        through radiative cooling of the bulk
        plasma~\cite{whytejournalofnuclearmaterials2003,granetznucl.fusion2006,hollmannnucl.fusion2013},
        and current quench mitigation via enhanced runaway electron
        current dissipation due to increased collision
        rates~\cite{granetznucl.fusion2006,paz-soldanphysicsofplasmas2014,boozerphys.plasmas2015}.
        The resulting plasma mixture of hydrogenic fuel, helium ash,
        and high-Z impurities, at a total atomic number density of
        $10^{20-21}$~m$^{-3},$ leads to a radiatively cooled bulk
        electron population, along with a minority runaway electron
        component at a number density of $10^{16-17}$~m$^{-3},$ which
        is sufficient to carry a plasma current of a few to 10
        mega-amperes (MA) in ITER discharges. This is a most unusual
        plasma that emphasizes a variety of radiative processes for
        plasma power loss, some combination of which point to the
        surprising role of minority relativistic runaway electrons.
        Although the presentation here focuses on the conditions of a
        disrupting tokamak plasma, the issues uncovered apply to a
        range of plasma applications in which a minority population of
        relativistic electrons co-exists with a cold thermal bulk that
        contains high-atomic-number ion species.  Some non-fusion
        examples include atmospheric
        lightning~\cite{gurevichphysicslettersa1992}, planetary
        radiation belts~\cite{fletcherspacescirev2011}, and any
        synchrotron emitting astrophysical plasma that is also metal
        rich~\cite{israelian-etal-book-2012}.
	
	In a mitigated tokamak disruption discharge where $n_{\rm
          Ar}=n_{\rm D}=10^{20}$~m$^{-3},$ line radiation by argon
        would dominate the plasma radiative power loss, which is
        around a few GW/m$^{3}$ over the electron temperature range of
        10~eV to 1~keV. For an ITER plasma of 300~m$^3$ in volume and
        $300$~MJ in thermal energy, this loss mechanism alone would
        ensure a thermal collapse (electron temperature, $T_e$, from a
        few keV to below 10~eV) over $1$~ms or less.  Once the
        electron thermal temperature reaches a few eV, radiative loss
        in a thermal plasma of comparable argon and deuterium number
        density becomes negligibly inefficient.  The minority runaway
        electron population, despite the tiny number density of
        $10^{16-17}$~m$^{-3}$ required to carry a few to 10 MA of
        current in ITER, would dominate the radiative loss.  The
        underlying cause for runaway electrons aiding efficient
        radiative power loss is fundamentally due to a quantum
        electrodynamics (QED) effect dating back to M{\o}ller, Breit,
        and Bethe.
	
	At electron energies significantly above the threshold energy,
        the excitation and ionization integrated cross-section (ICS) 
        would normally decrease with
        increasing impacting electron energy. However, this trend
        reverses when the impacting electron reaches relativistic
        energies, usually at a threshold around one MeV.  This is a
        lowest-order QED correction associated with the M{\o}ller or
        generalized Breit
        interaction~\cite{llovetjournalofphysicalandchemicalreferencedata2014,fontesj.phys.b:at.mol.opt.phys.2015,scofieldphys.rev.a1978,bostockphys.rev.a2013,fontesepjh2014}. Much
        of these physical effects are captured by the Bethe formula
        for electron stopping \textcolor{black}{~\cite{bethez.physik1932, inokutirev.mod.phys.1971,qedlifshitz,
          breizmannucl.fusion2019}
        \begin{align}
          -\frac{d{E}}{d x} = \sum_\alpha \frac{2 \pi e^4 n_{e\alpha}}{m_e v^2}
          \left[ \log \left( \frac{m_e^2 c^4 (\gamma^2 - 1)(\gamma - 1)}{
              2\left\langle I_{\alpha} \right\rangle^2} \right) \right . \notag
            \\ - \left . \left( \frac{2}{\gamma} - \frac{1}{\gamma^2}\right)
            \log 2 + \frac{1}{\gamma^2} + \frac{(\gamma - 1 )^2}{8\gamma^2}
            \right], \label{eq:bethe-stopping}
        \end{align}
        }where $c$ is the speed of light, $\epsilon_0$ the vacuum permittivity,
	$\beta=v/c,$ and $e$ and $m_e$ the electron charge and rest mass
	respectively. For each ion species in a specific charge state, which is
	denoted by subscript $\alpha$, there is a unique mean excitation
	energy $\left\langle I_{\alpha} \right\rangle$ and bound electron
	density $n_{e\alpha}.$ In fact, the standard treatment in runaway
	modeling is to incorporate the runaway slowing down due to excitation
	and ionization via a friction in the Fokker-Planck collision operator
	using the Bethe
	formula~\cite{hesslowj.plasmaphys.2018,hesslowphys.rev.lett.2017}.
	The effect of ion charge state distribution (CSD) on runaway and thermal
	bulk plasma evolution is through a collisional-radiative (CR) model that
	deals with only the background Maxwellian population at given
	temperature.
	
	In a dilute plasma such as those in tokamak disruption, the
	aforementioned decoupled treatment of background thermal electrons and
	runaways~\cite{hesslowj.plasmaphys.2018,hesslowphys.rev.lett.2017} has
	a straightforward prediction for the radiative power loss as the bulk
	electrons are cooled to below a few eV.  Namely, the
	background electron density drops precipitously due to cold thermal
	electron recombination with ions. Along with a mean charge number
	$\left< Z\right>$ approaching zero, the radiative power loss by
	thermal electrons can decrease by 5 orders of magnitude between
	$T_e=5$~eV to $T_e=1$~eV.  Meanwhile, as $\left\langle Z\right\rangle
	\rightarrow 0$ with $T_e\rightarrow 0,$ the simultaneous increase in
	bound electron density $n_{e\alpha}$ and decrease in mean excitation
	potential $\left\langle I_{\alpha} \right\rangle$ leads to enhanced
	radiative loss by runaways, according to the Bethe formula.
	
	Unlike in a solid, inelastic collisions between runaways and
        high-atomic-number impurities in a dilute plasma can directly
        alter the charge state of the plasma ions. The change in ion
        CSD, in turn, modifies both the radiative power loss by
        thermal electrons and the runaways. The former is through
        modified line emissions as the dominant bound-bound
        transitions have explicit charge state dependence, while the
        latter, using the Bethe stopping power formula, is by way of
        modified bound electron density $n_{e\alpha}$ and mean
        excitation potential $\left\langle I_{\alpha} \right\rangle$.
        This Letter provides CR modeling taking into account both the
        thermal bulk and the runaways with enhanced excitation and ionization 
        scattering processes, which allows us to not only
        elucidate the new qualitative trends enabled by the coupling
        between the minority runaways and bulk thermal population
        through high-atomic-number ions, but also quantify the
        discrepancies between a full CR treatment and a decoupled
        treatment of thermal bulk and minority runaways, which has
        been the state of the art in runaway modeling.
        \textcolor{black}{Supplementary material~\cite{garland-etal-prl-supplementary-2019},
          contrasts the inclusion of both excitation and ionization collision 
          with a recent analysis in the
          context of impurity transport that models the ionization
          balance using the Bethe stopping power formula for the
          runaway ionization rate~\cite{hollmannnucl.fusion2019}.}
	
	Our CR model is a fork (called FLYCHKLite) of the FLYCHK
        model/code, which is widely used in dense plasma
        applications~\cite{chunghighenergydensityphysics2005}, for its
        demonstrated applicability of the super-configuration model to
        sufficiently replicate the required physics in a
        computationally efficient
        manner~\cite{chunghighenergydensityphysics2013,pironhighenergydensityphysics2017}.
        The Los Alamos suite of atomic physics codes
        \cite{fontesj.phys.b:at.mol.opt.phys.2015} have been employed
        to calibrate and verify the model, especially the contribution
        from $\Delta n=0$ transitions in line emissions.  The
        robustness of our physics findings~\textcolor{black}{\cite{garland-etal-prl-Auger-comment-2019}} is further assessed via an
        uncertainty quantification analysis on the relativistic ICS
        for both excitation and ionization.
	
	{\em Relativistic inelastic scattering:} To account for the collisional
	processes of an arbitrary electron energy distribution function
	(EEDF), we need the ICS for electron
	impact excitation and ionization, in both the non-relativistic and
	relativistic energy regimes.  Our approach is to employ
	non-relativistic analytic fits used in FLYCHK, augmented by the
	addition of an analytic relativistic correction following a
	M{\o}ller-Bethe-like functional form
	\cite{bretagnej.phys.d:appl.phys.1986,scofieldphys.rev.a1978,fontesepjh2014}. For
	electron-impact excitation, the analytic form of van Regemorter
	\cite{vanregemorterastrophys.j.1962} is used for the non-relativistic
	regime,
	\begin{eqnarray}\label{collexc}
		\sigma_{i\rightarrow j}^{\text{NR}} = \frac{8\pi^2a_0^2}{\sqrt3} \left
		( \frac{Ry}{\Delta E_{ij}} \right)^2 \frac{f_{ij}g(U)}{U} ,
	\end{eqnarray}
	where $U=E/\Delta E_{ij}$ is the scaled incoming electron energy, 
	$Ry$ is the Rydberg energy, $f_{ij}$ is the collision oscillator
	strength and $a_0$ is the Bohr radius. The Gaunt factor is
	\begin{eqnarray} \label{gauntf}
		g(U) = A\log U + B + C(U+a)^{-1} + D(U+a)^{-2},
	\end{eqnarray}
	where coefficients are found via empirical fits of \eqref{gauntf} to averaged
	hydrogenic ICS computed by Chung \etal
	\cite{chunghighenergydensityphysics2007} via plane-wave Born calculations
	augmented by near-threshold scaling of Kim \cite{kimphys.rev.a2001}.
	
	Using \eqref{collexc} as a basis, a M{\o}ller-Bethe-like relativistic
	correction \cite{bretagnej.phys.d:appl.phys.1986,
		bethez.physik1932,mollerann.phys.1932} is applied so the total ICS
	is
	\begin{eqnarray}\label{exc_TOT}
		\sigma_{i\rightarrow j}^{\text{TOT}} = (1-S(E))\sigma_{i\rightarrow
			j}^{\text{NR}} + S(E)\sigma_{i\rightarrow j}^{\text{R}},
	\end{eqnarray}
	where
	\begin{eqnarray}\label{exc_R}
		\sigma_{i\rightarrow j}^{\text{R}} = \frac{8\pi a_0^2
			Ry^2}{m_ec^2\beta^2}\frac{f_{ij}}{\Delta E_{ij}} \left[ \log \left(
		\frac{\beta^2}{1-\beta^2}\frac{m_ec^2}{2\Delta E_{ij}} \right) -
		\beta^2 \right] ,
	\end{eqnarray}
	and the sigmoid smoothing function is defined as
	\begin{eqnarray}\label{sigmoid_ion}
		S(E) =  {\left[1+\exp(10^{-5}(10^5 - E))\right]}^{-1},
	\end{eqnarray}
	with $E$ in units of eV. The form of \eqref{sigmoid_ion} is chosen to
	adequately describe known relativistic ICS for inelastic processes.
	
	For electron impact ionization, the non-relativistic ICS is
	modeled by the semi-empirical expression of Burgess and Chidichimo
	\cite{burgessmon.not.r.astron.soc.1983}
	\begin{eqnarray} \label{burgession}
		\sigma_{(Z,i) \rightarrow (Z+1,j)}^{\text{NR}}= \pi a_0^2 C \xi \left(
		\frac{Ry}{\Delta I_Z^i}\right)^2 \frac{1}{U} \log(U) W(U),
	\end{eqnarray}
	where $\Delta I_Z^i$ is the ionization threshold energy for the level $i$ of
	charge state $Z$, $U=E/\Delta I_Z^i$, $\xi$ is the number of electrons in the
	shell being ionized, $C$ is an arbitrary constant nominally around 2
	\cite{burgessmon.not.r.astron.soc.1983}, and
	\begin{eqnarray} \label{burgessW}
		W(U) = (\log (U))^{\frac{\beta^*}{U}}, \\
		\beta^* = 0.25\sqrt{\frac{100Q_n + 91}{4Q_n + 3}} - 1.25,
	\end{eqnarray}
	where $Q_n$ is the screened charge \cite{chunghighenergydensityphysics2005}.
	
\begin{figure}
		\centering
		\includegraphics[width=\linewidth]{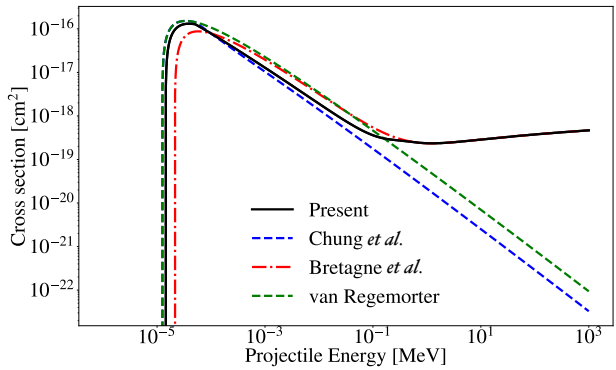}
		\caption{ICS for excitation of Cl-like Ar$^+$ $n=3 \rightarrow n=4$ transition.
			Comparison of our proposed approximate ICS is made against the non-relativistic
			results of Chung \textit{et al.} \cite{chunghighenergydensityphysics2007}, van
			Regemorter \cite{vanregemorterastrophys.j.1962} and the relativistic result of
			Bretagne \textit{et al.} \cite{bretagnej.phys.d:appl.phys.1986}.}
		\label{fig:ar-exc}
	\end{figure}

\begin{figure}
		\centering
		\includegraphics[width=\linewidth]{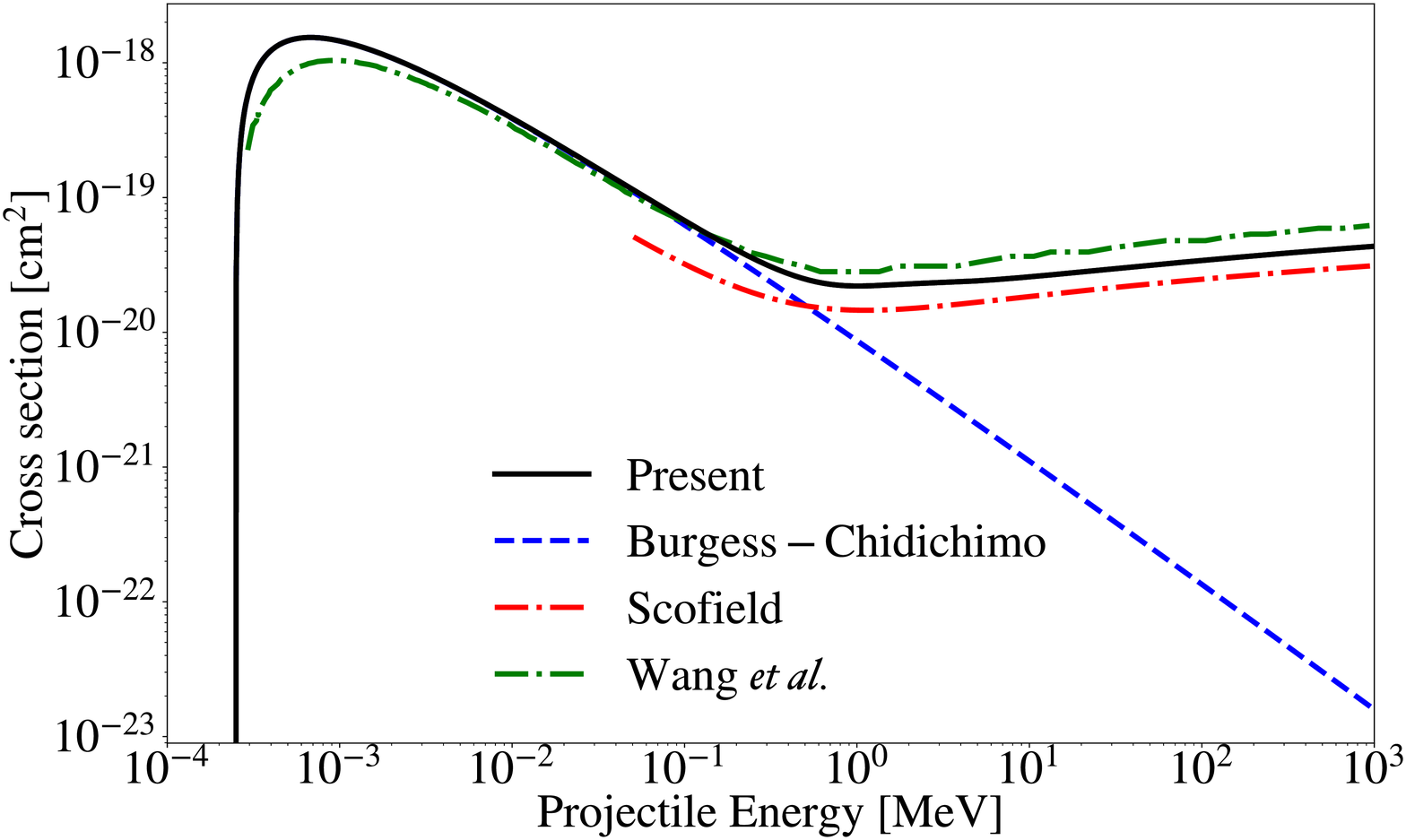}
		\caption{ICS for lumped-average L-shell ionization of Ar. Comparison of our
			proposed approximate ICS is made against the non-relativistic result of Burgess
			and Chidichimo \cite{burgessmon.not.r.astron.soc.1983}, and the relativistic
			results of Scofield \cite{scofieldphys.rev.a1978} and Wang \textit{et al}.
			\cite{wangj.phys.bat.mol.opt.phys.2018}.}
		\label{fig:ar-ion}
	\end{figure}

	Again using the same form as \eqref{exc_TOT} for the total
        ionization ICS, we add a relativistic correction
        \cite{scofieldphys.rev.a1978,bethe1933} on top of
        \eqref{burgession} following the form
	\begin{align}
		\sigma_{(Z,i) \rightarrow (Z+1,j)}^{\text{R}} = & \pi a_0^2 C \xi  \left(\frac{Ry}{\Delta I_Z^i} \right) \nonumber \\
		& \left[  \log\left( \frac{\beta^2}{1-\beta^2} \frac{m_ec^2}{2\Delta I_Z^i} \right)  - \beta^2  \right].\label{ionizgeneral}
	\end{align}
	Figs.~\ref{fig:ar-exc} and~\ref{fig:ar-ion} compare the
        proposed fits with calculated ICS for excitation and
        ionization in the literature. We note agreement within a
        factor of two of recent R-matrix theories at the peak of the
        cross section, and that unlike those reported in
        Ref.~\cite{wangj.phys.bat.mol.opt.phys.2018}, normally such
        calculations are not extended into the relativistic regime of
        relevance here.	\textcolor{black}{The plasma physics impact established in this paper
        certainly points to the importance and urgency of such
        systematic first-principle cross section calculations.}
	
        {\em Simulation problem setup:} To demonstrate the impact of
        using ICS that describe enhanced scattering observed at
        relativistic energies we perform a comparison between CR model
        results computed with and without relativistic ICS effects
        applied to conditions representative of a tokamak disruption
        mitigation scenario. Without losing generality, the electron
        distribution model is defined as a bulk Maxwellian at given
        temperature $T_e$ and a Gaussian runaway tail that has a mean
        energy $\left<E_0\right>$ and full width at half maximum
        spread of $\Delta E_{\rm FWHM},$ at runaway number density
        $n_{\rm RE}.$ The ion population is fixed with $(n_{\rm D},
        n_{\rm Ar}).$ The thermal bulk then has an electron density
        ($n_e$) from the quasi-neutral condition $ n_e = \left\langle
        Z\right\rangle_{\rm Ar} n_{\rm Ar} + \left\langle
        Z\right\rangle_{\rm D} n_{\rm D} - n_{\rm RE}.$ As $T_e$
        varies from 10~keV to 1~eV, $n_e$ can vary greatly as a
        function of $\left\langle Z\right\rangle$, but $n_{\rm RE}$ is
        a fixed number that only depends on how much current the runaways
        need to carry. For example, to carry $10$~MA of runaway current in ITER,
        $n_{\rm RE}\approx 10^{17}$m$^{-3}.$ Our scan over $T_e$ is
        performed with fixed $\left<E_0\right>= 10 \ \text{MeV},
        \Delta E_{\rm FWHM}= 5 \ \text{MeV}, n_{\rm RE}= 10^{17}
        \ \text{m}^{-3},$ and $n_{\rm D}=n_{\rm Ar}=10^{20}
        \ \text{m}^{-3}.$ The CR model iteratively solves for
        $\left\langle Z\right\rangle$ and $n_e$ when everything else
        is fixed.
	
\begin{figure}
		\centering
		\includegraphics[width=\linewidth]{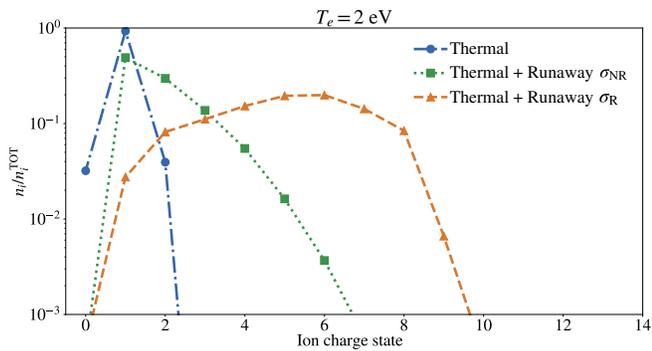}
		\caption{Argon ion CSD
			at $T_e = 2$ eV with a relativistic Gaussian tail supporting 
			a 10 MA current compared to a thermal discharge. The impact of 
			including relativistic ICS effects ($\blacktriangle$) compared to using ICS 
			without the M{\o}ller/generalized Breit interaction ($\blacksquare$) or simply assuming
			a thermal plasma ($\bullet$) is apparent.}
		\label{fig:ar-Z_states}
	\end{figure}

	{\em Runaway enhanced ionization at low $T_e$:} With the
        addition of a relativistic tail, and the inclusion of enhanced
        inelastic scattering by this relativistic electron tail, a
        clear increase in $\left\langle Z\right\rangle$ and broadening
        of the CSD is observed for the argon discharge at low $T_e$,
        exemplified by Fig. \ref{fig:ar-Z_states} for $T_e=2$ eV.
        Here, we see the addition of relativistic electrons sampling
        {\em non-enhanced} ICS increases $\left\langle Z\right\rangle$
        slightly, and then the addition of {\em enhanced relativistic}
        ICS to the model further increases $\left\langle
        Z\right\rangle$ and promotes CSD broadening. This is a direct
        result of the relativistic tail of the EEDF being able to
        sample an increasing ICS at higher energies. A wider spread of
        charge states shown in Fig.  \ref{fig:ar-Z_states} will also
        lead to very different radiative cooling pathways, discussed
        in the next section. \textcolor{black}{Contrasting the yellow
          line with blue and green lines in
          Fig.~\ref{fig:ar-Z_states}, one can also see that the complex
          excitation-ionization pathways enabled by QED enhancement of
          the cross sections can deplete the Ar$^+$ population.  }
          
\begin{figure}
		\centering
		\includegraphics[width=0.85\linewidth]{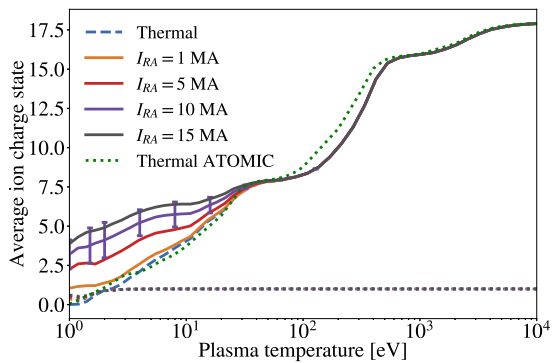}
		\caption{Average argon charge state, $\left\langle
			Z\right\rangle$, over thermal $T_e$ demonstrating
			enhancement produced by relativistic electrons sampling
			enhanced relativistic ICS. $\left\langle
			Z\right\rangle$ for a thermal plasma computed by LANL ATOMIC
			code \cite{fontesj.phys.b:at.mol.opt.phys.2015} shown for
			benchmarking purposes. Average
			charge state of deuterium (\textellipsis) shown for reference,
			indicating a fully stripped ion for nearly all plasma
			temperatures considered.}
		\label{fig:ar-Z-bar}
	\end{figure}

	Considering $\left\langle Z\right\rangle$ variation with $T_e$ Fig.
	\ref{fig:ar-Z-bar} demonstrates a clear increase in $\left\langle
	Z\right\rangle$ at $T_e < 50$ eV, when compared to a Maxwellian EEDF
	typically used in decoupled CR modeling. For benchmarking purposes the result
	for a thermal plasma computed with the LANL CR modeling code, ATOMIC
	\cite{fontesj.phys.b:at.mol.opt.phys.2015}, is included in Fig.
	\ref{fig:ar-Z-bar}, and agrees with the present thermal model.	
	
        {\em Runaway-induced ion charge state effects:} The
        runaway-induced ionization at low $T_e,$ which is expected for
        the current quench phase, can impact disruption physics and
        its modeling in a number of ways.  These are generally known
        as ion charge state effects, and here we give a few most
        prominent examples.  The first is the enhanced pitch angle
        scattering of runaways due to the partial screening effect,
        which has a direct dependence on the ion charge state
        distribution, according to Hesslow \textit{et
          al.}~\cite{hesslowphys.rev.lett.2017,hesslowplasmaphys.control.fusion2018}.
        The physical importance is that enhanced pitch angle
        scattering would limit the energy of the O-point of the
        runaway vortex, which is responsible for a lower runaway
        energy but broader pitch
        distribution~\cite{mcdevittplasmaphys.control.fusion2018,guoplasmaphys.control.fusion2017}.
        The runaway-induced ionization would then modulate the pitch
        angle scattering rate due to partial screening, impacting both
        the avalanche threshold and runaway energy distribution, as
        well as the spatial transport of the
        runaways~\cite{McDevitt_2019}.
	
\begin{figure}
		\centering
		\includegraphics[width=0.85\linewidth]{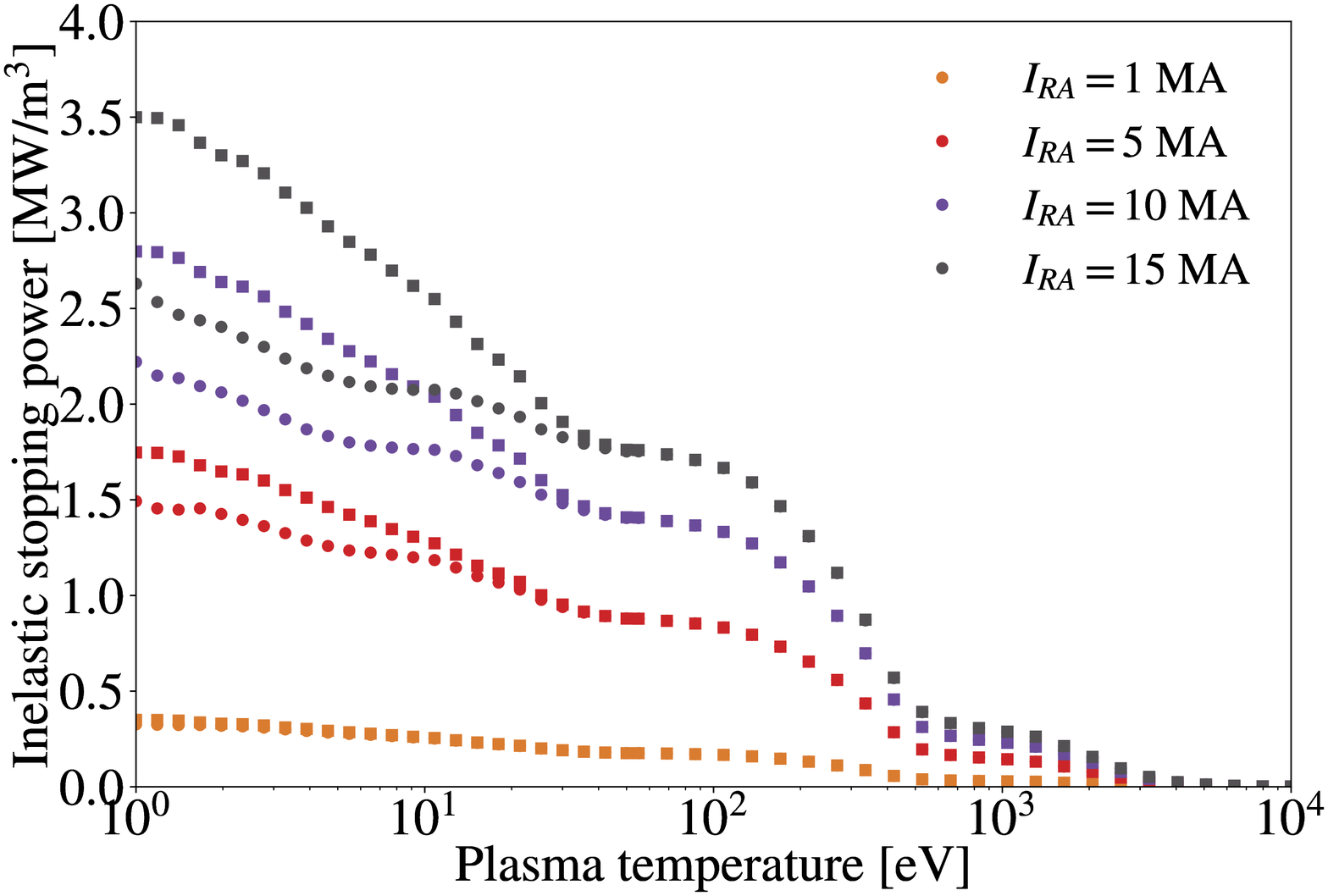}
		\caption{Inelastic collision stopping power experienced by
			relativistic runaway electrons colliding with argon
			ions. Values computed with the Bethe stopping power formula
			Eq.~(\ref{eq:bethe-stopping}) demonstrate the impact of coupling a runaway-enhanced
			CSD ($\bullet$) compared to using a thermal plasma CSD ($\blacksquare$). }
		\label{fig:ar-stopping}
	\end{figure}

	The second example is the slowing down of the runaways by inelastic
	collisions with high-atomic-number impurities. The runaway-induced ion
	charge state effect can be seen from the ICS for both excitation and
	ionization, Eqs.~(\ref{exc_R}) and (\ref{ionizgeneral}).  Both ICS tend to
	decrease with higher ion charge number since the excitation energy
	$\Delta E_{ij}$ and ionization threshold energy $\Delta I_Z^i$ are
	generally higher with more electrons stripped from the atom.  In the
	case of ionization ICS, the bound electron density $n_{e\alpha}$ would
	also decrease linearly with higher ion charge number.  The impact of  
	decreasing ICS with higher charge state 
	is demonstrated by contrasting the Bethe stopping power. 
	In Fig.~\ref{fig:ar-stopping}, we
	contrast the Bethe stopping power using (i) a CSD
	from the CR prediction of a background Maxwellian alone,
	and (ii) CSD from the CR prediction that
	includes the runaway contribution. There is significant over-estimate
	of the Bethe stopping power if the runaway-induced ionization is not
	taken into account at low $T_e$, which is crucial in a post-thermal-quench plasma. 	
	
        The third example is more conventional, and it concerns the effect of
	Coulomb collisions on thermal bulk electrons. This comes from the
	strong $Z_{\rm eff}$ dependence for thermal plasma conductivity, and
	the collisional damping of both the externally
	injected~\cite{guophys.plasmas2018} and
	self-induced~\cite{liuphys.rev.lett.2018} fast waves, by thermal
	electrons. The runaway-induced ionization, in the plasma regime of a
	mitigated tokamak disruption, can dramatically increase $Z_{\rm eff},$
	and hence increase dissipation of the remnant Ohmic current and the
	fast waves. The latter is particularly unfortunate since resonant
	wave-particle interaction offers a valuable tool for controlling the
	runaway energy~\cite{guophys.plasmas2018}.
	
\begin{figure}
		\centering
		\includegraphics[width=0.85\linewidth]{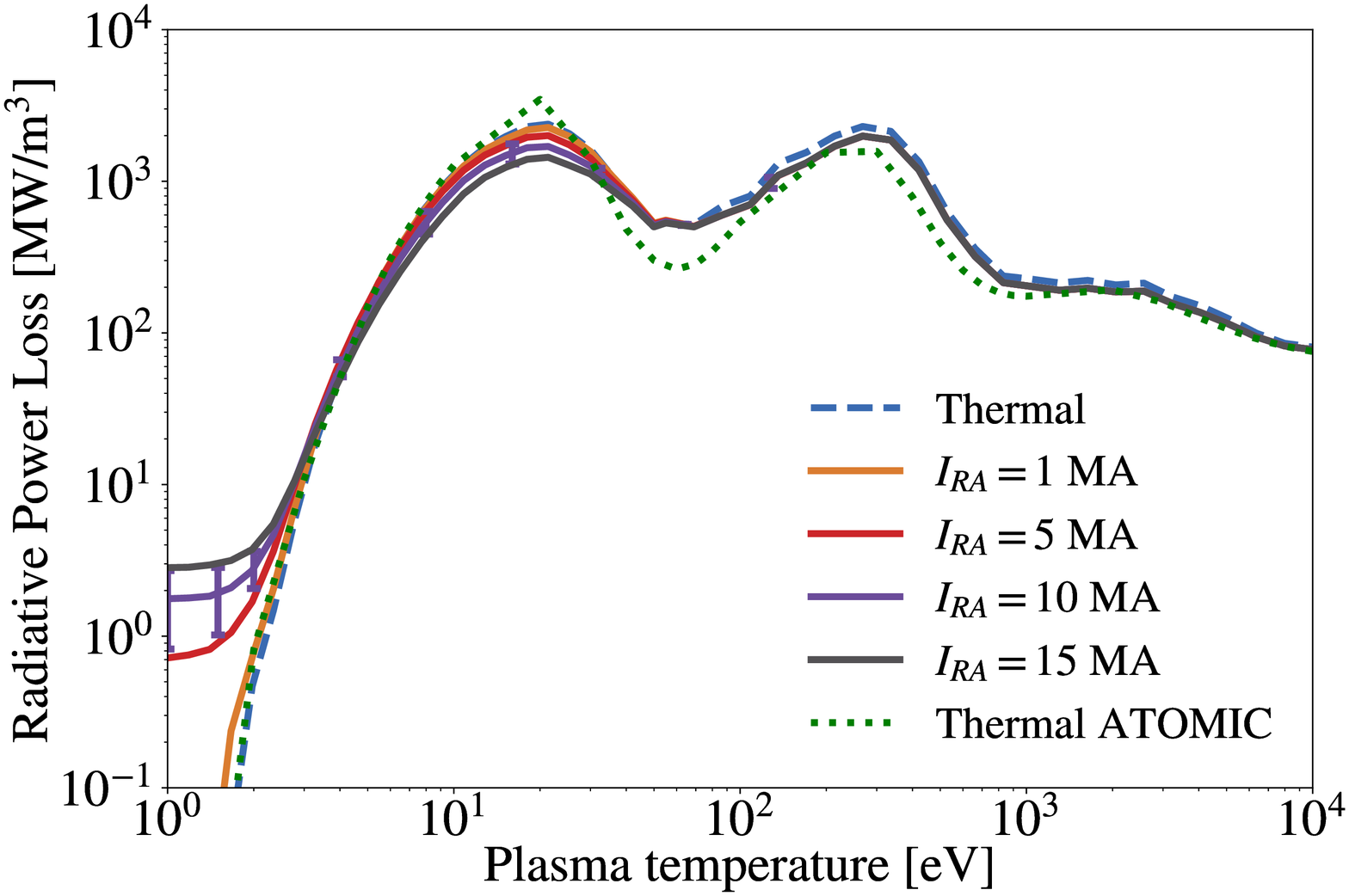}
		\caption{RPL of argon over thermal
			$T_e$ demonstrating enhanced RPL produced by relativistic electrons
			sampling relativistic inelastic ICS. RPL of thermal plasma computed with 
			LANL ATOMIC code \cite{fontesj.phys.b:at.mol.opt.phys.2015} is shown for benchmarking purposes.  }
		\label{fig:ar-RPL}
	\end{figure}

	{\em Runaway-induced radiative power loss (RPL) effects:} While it is
	expected that radiative power loss, shown in Fig.~\ref{fig:ar-RPL}, would be
	dominated by collisional excitation and ionization by runaways at low $T_e$ ($T_e<4$~eV), by virtue
	of a higher charge state,
	there is also a subtler runaway effect when RPL
	rates by thermal electrons and runaways are comparable.  This can be
	seen in Fig. \ref{fig:ar-RPL}, in the range of $5\leq T_e \leq 100$
	eV, where the RPL of the combined thermal plus runaway plasma is
	up to 40\% lower than that of the pure thermal plasma. 
	The reduced RPL of the runaway enhanced discharges 
	is found to be a result of the relativistic ICS
	enhancement diffusing and decreasing both the relative population of ion charge states and
	excited levels, resulting in a smaller RPL in this temperature range.	
	At higher temperatures, all RPL results effectively converge above
	approximately 100 eV, which is also seen in the result of the average
	charge state. Convergence occurs once the thermal population begins to
	dominate the ionization balance of the discharge, negating the
	excitation and ionization channels introduced by the relativistically
	enhanced scattering.
	
	{\em Uncertainty quantification:} To validate the qualitative
	trends observed in this study, we employed UQ software
	Dakota \cite{Dakota} to perform an epistemic global interval estimation 
	uncertainty analysis with
	relativistic electron ICS. In this UQ analysis, prefactors,
	$0.5\leq C_E,C_I \leq 2$, were applied to the relativistic excitation and
	ionization ICS formulas \eqref{exc_R} and
	\eqref{ionizgeneral} to describe uncertainty in the approximate formulas
	used in this work. For $\left\langle
	Z\right\rangle$ and RPL appearing in Figs.  \ref{fig:ar-Z-bar} and
	\ref{fig:ar-RPL}, uncertainty bounds are shown. The confidence intervals computed
	via 500 model evaluations with Latin hypercube sampling 
	over the range of ICS uncertainty supports our
	observation that there is a definitive enhancement in both
	$\left\langle Z\right\rangle$ and total RPL of argon ions considered
	as a direct result of relativistic electrons sampling relativistically
	enhanced ICS. 
	
	{\em Conclusion:} A minority relativistic electron component, at a
	number density three orders of magnitude smaller than the
	low-temperature thermal bulk, can dominate the CSD
	in the presence of high-atomic-number impurities, due to
	the relativistic enhancement of the collisional excitation and
	ionization ICS. The resulting charge state effects can
	impact (1) radiative power loss rate and related Bethe stopping power
	on the runaways, and (2) collisional dissipation of Ohmic current and
	externally-injected and self-excited plasma waves. These physics
	findings suggest the necessity of a coupled CR model that
	takes into account both the bulk electron population and the minority
	relativistic electron component, if high-atomic-number impurities are
	present.
	
	This work was jointly supported by the U.S. Department of
        Energy through the Magnetic Fusion Theory Program and the
        Tokamak Disruption Simulation (TDS) SciDAC project at both Los
        Alamos National Laboratory (Contract No. 89233218CNA000001)
        and Sandia National Laboratory (Contract No. DE-NA-0003525).
        Additional support is provided by Laboratory Directed Research
        and Development program of Los Alamos National Laboratory
        under project number 20200356ER.  T.~Elder was also supported
        by the Science Undergraduate Laboratory Internships (SULI)
        program.

	
	\bibliographystyle{apsrev4-1}

\end{document}